\documentclass[conference]{IEEEtran}
\IEEEoverridecommandlockouts
\usepackage{cite}
\usepackage{amsmath,amssymb,amsfonts}
\usepackage{graphicx}
\usepackage{textcomp}
\usepackage{xcolor}
\usepackage{listings}
\usepackage{algorithm}
\usepackage{algpseudocode}
\usepackage{comment}
\usepackage{hyperref}
\usepackage{lstautogobble}
\usepackage{subcaption}
\usepackage{booktabs}
\usepackage{tikz}
\usepackage{stfloats}
\usepackage{enumitem}
\setlist{nolistsep}
\usepackage[normalem]{ulem}
\usepackage{tcolorbox}
\usepackage{tabularx}
\usepackage{siunitx}

\def\BibTeX{{\rm B\kern-.05em{\sc i\kern-.025em b}\kern-.08em
    T\kern-.1667em\lower.7ex\hbox{E}\kern-.125emX}}

\newcommand*\circled[1]{\tikz[baseline=(char.base)]{
            \node[shape=circle,draw,inner sep=1pt] (char) {\textcolor{black}{#1}};}}

\makeatletter
\DeclareUrlCommand\ULurl@@{%
  \def\UrlLeft{\uline\bgroup}%
  \def\UrlRight{\egroup}}
\def\ULurl@#1{\hyper@linkurl{\ULurl@@{#1}}{#1}}
\DeclareRobustCommand*\ULurl{\hyper@normalise\ULurl@}
\makeatother

\newcommand{\kefah}[1] {{\color{blue}{Kefah}}: {\color{red}{#1}}}

\newcommand{\edit}[1] {{#1}}

\newcommand{\tool}[0]{\mbox{\textsc{ucTrace}}}

\makeatletter
\newif\if@ucxfirst \@ucxfirsttrue
\newif\if@uctfirst \@uctfirsttrue
\newif\if@ucpfirst \@ucpfirsttrue

\newcommand{\ucx}{%
    \if@ucxfirst
        Unified Communication X (UCX)%
        \global\@ucxfirstfalse
    \else
        UCX%
    \fi
}

\newcommand{\uct}{%
    \if@uctfirst
        UCT (UCX Transport)%
        \global\@uctfirstfalse
    \else
        UCT%
    \fi
}

\newcommand{\ucp}{%
    \if@ucpfirst
        UCP (UCX Protocol)%
        \global\@ucpfirstfalse
    \else
        UCP%
    \fi
}

\makeatother
\newcommand{\am}[0]{{active message}}
\newcommand{\ep}[0]{{endpoint}}
\newcommand{\md}[0]{{memory domain}}
\newcommand{\iface}[0]{{interface}}

\newcommand{\sanitizer}[0]{{NVIDIA Compute Sanitizer API}}
\newcommand{\nvidia}[0]{{NVIDIA}}

\definecolor{bluekeywords}{rgb}{0.13, 0.13, 1}
\definecolor{greencomments}{rgb}{0, 0.5, 0}
\definecolor{redstrings}{rgb}{0.9, 0, 0}
\definecolor{graynumbers}{rgb}{0.5, 0.5, 0.5}
\definecolor{background_ok}{rgb}{0.95, 0.95, 0.95}

\begin{document}

\title{\tool{}: A Multi-Layer  Profiling Tool for UCX-driven Communication}

\author{\IEEEauthorblockN{Emir Gencer}
\IEEEauthorblockA{\textit{Koç University} \\
Turkey \\
egencer20@ku.edu.tr}
\and
\IEEEauthorblockN{Mohammad Kefah Taha Issa}
\IEEEauthorblockA{\textit{Koç University} \\
Turkey \\
missa18@ku.edu.tr}
\and
\IEEEauthorblockN{Ilyas Turimbetov}
\IEEEauthorblockA{\textit{Koç University} \\
Turkey \\
iturimbetov18@ku.edu.tr}
\and
\IEEEauthorblockN{James D. Trotter}
\IEEEauthorblockA{\textit{Simula Research Laboratory} \\
Norway \\
james@simula.no}
\and
\IEEEauthorblockN{Didem Unat}
\IEEEauthorblockA{\textit{Koç University} \\
Turkey \\
dunat@ku.edu.tr}
}
\maketitle

\begin{abstract}
\ucx{} is a communication framework that enables low-latency, high-bandwidth communication in HPC systems. With its unified API, \ucx{} facilitates efficient data transfers across multi-node CPU-GPU clusters. 
\ucx{} is widely used as the transport layer for MPI, particularly in GPU-aware implementations. However, existing profiling tools lack fine-grained communication traces at the \ucx{} level, do not capture transport-layer behavior, or are limited to specific MPI implementations.


To address these gaps, we introduce \tool{}, a novel profiler that exposes and visualizes \ucx{}-driven communication in HPC environments. \tool{} provides insights into MPI workflows by profiling message passing at the \ucx{} level, linking operations between hosts and devices (e.g., GPUs and NICs) directly to their originating MPI functions. Through interactive visualizations of process- and device-specific interactions, \tool{} helps system administrators, library and  application developers optimize performance and debug communication patterns in large-scale workloads. We demonstrate \tool{}'s features through wide range of experiments 
including MPI point-to-point behavior under different \ucx{} settings, {\tt Allreduce} comparisons across MPI libraries, communication analysis of a linear solver, NUMA binding effects, and profiling of GROMACS MD simulations  with GPU acceleration at scale. 
\tool{} is publicly available at \url{https://github.com/ParCoreLab/ucTrace}. 
\end{abstract}


\renewcommand{\sout}[1]{}

\section{Introduction}
\label{sec:intro}
Unified Communication X (UCX) is a high-performance communication framework designed to support low-latency, high-bandwidth data transfers in parallel applications \cite{ucx, ucx-paper}. It improves message-passing performance in several widely used frameworks, such as Open MPI~\cite{openmpi}, MPICH~\cite{mpich}, OpenSHMEM \cite{openshmem, openshmem-paper}, NCCL\footnote{NCCL can use UCX as underlying communication layer, when the UCX plugin is used (\url{https://github.com/Mellanox/nccl-rdma-sharp-plugins})}, and NVSHMEM\footnote{NVSHMEM can be configured to use \ucx{} with an environment variable.}, and enables scalable execution across both CPU- and GPU-based systems. By providing a unified API that abstracts multiple transport technologies, including InfiniBand, RDMA, TCP, and GPUDirect \cite{gpudirect}, UCX allows applications to communicate efficiently without being tied to a specific interconnect. A notable use case is CUDA-aware MPI, where UCX eliminates the need for explicit host memory copies by allowing GPU memory pointers to be passed directly to MPI functions~\cite{cuda-aware-mpi}.

Profiling communication patterns and identifying scalability bottlenecks are critical tasks for achieving high performance on modern HPC systems. Understanding how communication unfolds within applications across CPUs, GPUs, and network interfaces is essential. A detailed view of the communication stack, including the data paths, protocols, transport mechanisms, communication topology, and the underlying hardware resources, can significantly enhance both application and system efficiency. System administrators, communication library and application developers all benefit from comprehensive UCX profiling tools that support performance tuning and provide actionable insights into communication behavior.

Several tools have been developed to support profiling and visualization of communication in multi-node and GPU-accelerated systems. Tools such as Extrae~\cite{extrae} with Paraver~\cite{paraver}, OSU INAM~\cite{Inam-Cross-Stack}, Nsight Systems~\cite{nsight-systems}, and Snoopie~\cite{Snoopie} offer varying levels of support for communication profiling. While Extrae and Paraver provide detailed tracing and flexible visualizations, they lack visibility into transport-layer behavior. OSU INAM collects network metrics via an InfiniBand daemon and combines them with MPI statistics, but it is specifically designed for MVAPICH, relies on non-standard MPI\_T extensions, and requires a system-wide daemon. NVIDIA’s Nsight Systems can trace MPI and some \ucx{} calls but provides limited visibility into communication directionality. Snoopie focuses on intra-node GPU communication and lacks support for CUDA-aware MPI. To our knowledge, none of these tools offers profiling capabilities with a specific focus on \ucx{}.

To address these limitations, we introduce \tool{}, a multi-node profiling and visualization tool that captures and analyzes \ucx{} communication activity. \tool{} effectively profiles applications that use \ucx{} as their transport layer, including both standard and GPU-aware MPI implementations. It provides a multi-layer view of communication behavior, enabling developers to trace interactions from the low-level UCX transport layer, through protocol-level operations, and up to MPI functions. This layered analysis supports the needs of diverse users: system administrators can validate library configurations, library developers can investigate protocol behavior and transport usage, and application developers can gain insights into communication patterns involving CPUs, GPUs, and NICs via an interactive visualization interface equipped with various filtering options.

We demonstrate the utility of \tool{} through several experiments with both CPU and GPU applications on MareNostrum and Leonardo supercomputers. These include analyses of eager and rendezvous protocol usage at the UCX level, comparison of MPI {\tt Allreduce} algorithms employed by Open MPI and MPICH, visualization of communication graphs in a conjugate gradient solver~\cite{acg} and a molecular dynamics simulation (GROMACS~\cite{gromacs}), and detection of performance bugs related to task placement. 
These case studies demonstrate how \tool{} can be used across different users, application domains and abstraction levels. Our contributions are:

\begin{itemize}
    \item A profiling and analysis tool for tracking \ucx{} communication across distributed memory systems. 
    \item Mapping of \ucx{} events to corresponding MPI-level operations for improved interpretability. 
    \item GPU device attribution for CUDA-aware MPI, enabling detailed tracing of GPU-originated communication.
    \item A flexible visualization interface for exploring communication patterns at process, device, and transport levels. 
    \item Extensive case studies demonstrating \tool{}’s utility in diverse scenarios 
\end{itemize}

Although \tool{} may not be as comprehensive as other tools that have been around for 10+ years;  it sets itself apart by addressing an emerging need through its multi-layer communication attribution for \ucx{}. \tool{} links MPI communication to transport-layer events in the UCX stack, visualizes interactions across the GPU, NIC, and host without requiring privileged setup or being tied to a specific MPI implementation. While \tool{} currently focuses on MPI and CUDA-aware MPI applications, its reliance on the UCX transport layer enables future support for NVSHMEM, NCCL, and ROCm-aware MPI \cite{rocm-aware-MPI} with modest extensions.

\section{Background and Related Work}
\label{sec:background}

 \ucx{} is structured into three layers: \ucp{}, which offers high-level communication primitives such as messaging and remote memory access (RMA); \uct{}, which provides low-level access to network hardware; and UCS (Support), which includes utilities for memory management and threading. 

\begin{figure*}
\begin{minipage}[c]{0.27\textwidth}
\includegraphics[width=0.94\textwidth]{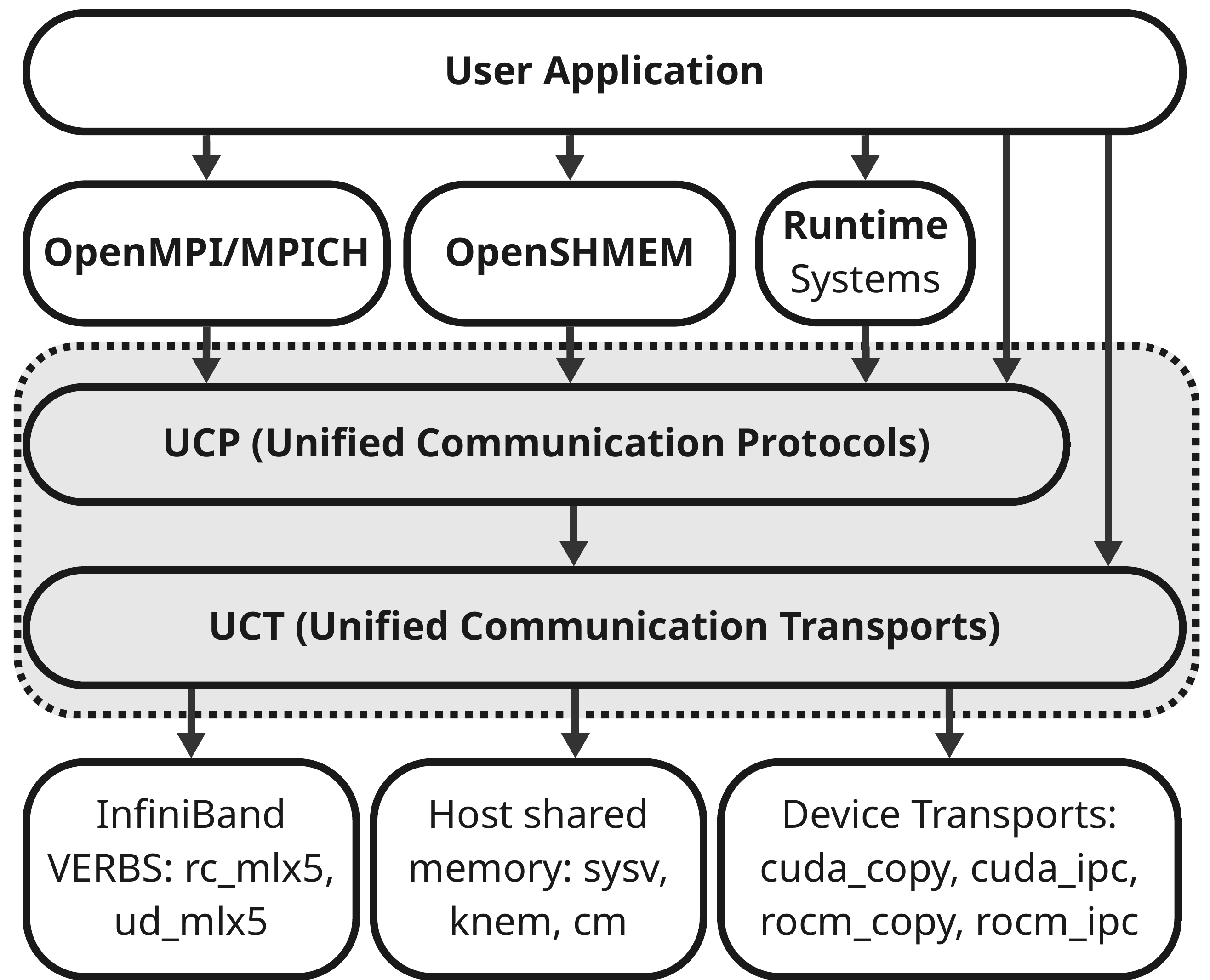}
    \caption{\ucx{} architecture}
    \label{fig:ucx-stack}
\end{minipage}
\hfill
\begin{minipage}[c]{0.67\textwidth}
\includegraphics[width=\textwidth]{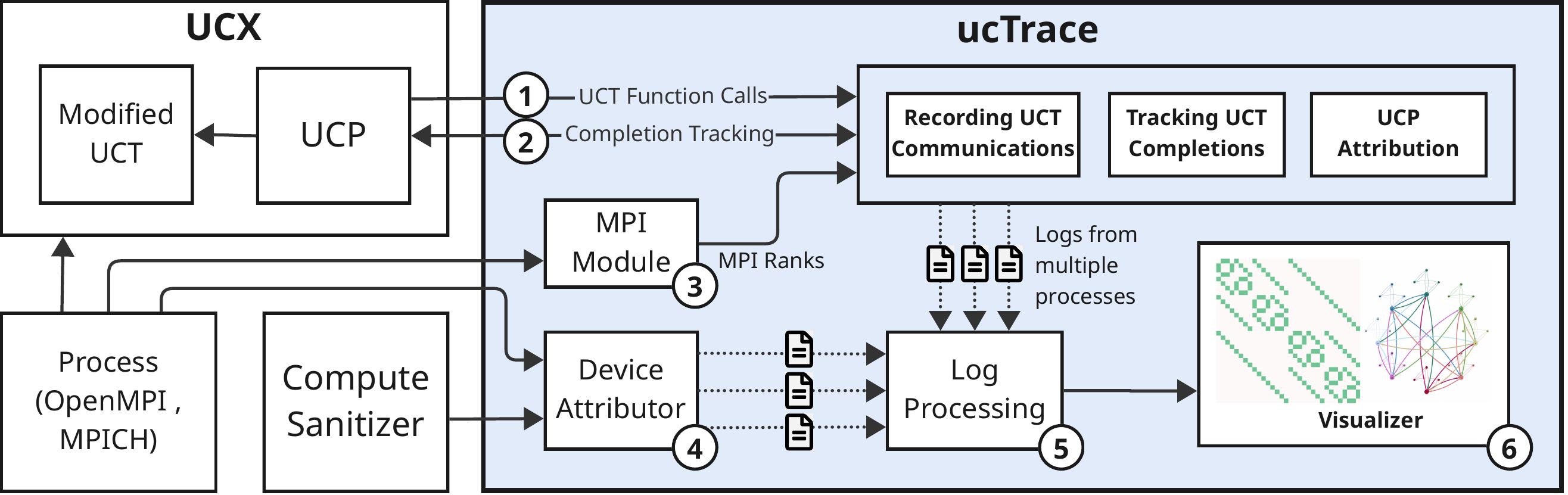}
    \caption{Overview of \tool{} and its components.}
    \label{fig:tool-flowchart}
\end{minipage}
\end{figure*}

Figure \ref{fig:ucx-stack} displays how user applications and communication libraries interact with \ucx{} stack. \uct{} supports various transports leveraging InfiniBand VERBs for inter-node communication and host shared memory, CUDA and ROCm functionalities for intra-node communication. During runtime, a \ucp{} communication dynamically selects \uct{} transports based on the passed pointer type and system topology. For instance, when a CUDA pointer is passed to a inter-node \ucp{} communication, the {\tt{rc\_mlx5}} transport might be utilized, whereas for a same-node communication, the {\tt{cuda\_ipc}} transport is preferred. Consequently, \ucp{} can accept device pointers as data payloads and transfer data using the appropriate mechanism, making it inherently GPU-aware. This capability enables higher-level communication libraries, including Open MPI and MPICH, to support device pointers in their communication primitives, thereby extending GPU-awareness to these libraries. 

OpenFabrics Interface (OFI)~\cite{libfabric} is another communication framework enabling scalable network communication in HPC systems. Although popular MPI implementations like Open MPI and MPICH support both \ucx{} and OFI, we focus on \ucx{} due to its higher-level API and stronger GPU communication integration.


\subsection{Related work}
\label{sec:related-work}

With the continuing rise of multi-node communication frameworks and technologies \cite{unat2024landscapegpucentriccommunication}, there exist several tools that offer communication profiling and visualization capabilities.  

Given MPI’s widespread use, many tools offer extensive profiling support. mpiP \cite{mpip} is a lightweight library that collects statistical data on MPI call execution times and can be extended to generate communication matrices \cite{mpip-comm-matrix}. 
EZTrace \cite{ez-trace} is a generic performance analysis tool that can generate  communication matrices of the MPI application.
Extrae \cite{extrae} is an instrumentation tool to trace parallel programs that use MPI or OpenMP and supports GPU kernel profiling with CUDA and CUPTI~\cite{nvidia-cupti}. It 
supports hardware counters through PAPI \cite{PAPI}, and incorporates advanced analysis features such as structure detection and periodicity analysis \cite{extrae-ml}. Its traces are visualized using Paraver~\cite{paraver}, which supports flexible timelines, tables, and histograms.

The Scalasca toolset~\cite{scalasca,scalasca-v2} is a widely adopted suite for profiling parallel applications. It relies on Score-P\cite{score-p} for instrumentation, the successor of VampirTrace \cite{vampir}, and supports automatic performance analysis for MPI and OpenMP programs. It presents data using metric, call tree, and system topology views, and is extensible via plugins \cite{15years-scalasca-paraver}. Traces generated by Scalasca can also be visualized in Vampir~\cite{vampir}, enabling the generation of communication matrices.

While powerful, these tools do not expose transport-layer communication details or NIC-level activity. They lack visibility into the underlying transport mechanisms used by UCX and do not directly attribute communication to specific NICs, as \tool{} does.

The authors of~\cite{PERUSE} showed that multiple data transfers within a single non-blocking MPI call could be visualized with peer task send/receive timings to infer network performance. This was achieved by implementing the PERUSE specification~\cite{mpi-peruse-spec} in Open MPI, collecting metrics via PERUSE events, and visualizing them using Paraver \cite{paraver}. However, performance concerns with PERUSE led to the introduction of the MPI\_T events interface~\cite{MPI-T} as an alternative.  A notable tool leveraging MPI\_T is OSU INAM (InfiniBand Network Analysis and Monitoring)~\cite{Inam-Cross-Stack}.

OSU INAM is arguably the closest related tool to \tool{}. It collects network metrics via an InfiniBand daemon, gathers MPI metrics through the MPI\_T interface, and correlates the two to provide insights such as real-time link utilization, network topology, and MPI primitive usage. INAM was later extended with CUPTI support to incorporate CUDA information~\cite{Inam-GPU}. However, it relies on proposed MPI\_T extensions that are not yet part of the standard~\cite{Inam-Cross-Stack}, and its MPI integration is primarily designed and tested with MVAPICH. 
Additionally, INAM requires a system-wide daemon initiated by an administrator~\cite{Inam-1}, which reduces accessibility for end users. In contrast, \tool{} operates at the \ucx{} layer, supports multiple MPI implementations and potentially other communication libraries, and is more accessible without requiring privileged setup.

When comes to GPU-specific tools, ComScribe\cite{ComScribe}, one of the earliest tools targeting GPU communication and focuses on monitoring NVLink traffic and NCCL-based applications. 
Snoopie\cite{Snoopie} supports direct remote memory accesses between GPUs within a node. While it supports NCCL and NVSHMEM, it lacks inter-node communication support, particularly for CUDA-aware MPI. \tool{} builds on Snoopie's approach for device attribution in CUDA-aware MPI.
Nsight Systems~\cite{nsight-systems}, NVIDIA's system-wide profiler, traces MPI and some \ucx{} protocol layer  calls, providing call types, durations, and ordering. However, it does not capture the directionality of communication beyond the basic one-to-one calls or offer the same level of visualization as \tool{}.

In summary, \tool{} distinguishes itself through multi-layer attribution of communication: it links high-level MPI calls to low-level transport events at the \uct{} layer, attributes communication to \ucp{} events and GPU devices, captures directionality, and visualizes interactions across GPU, NIC, and host components. Operating at the \uct{} layer, \tool{} is agnostic to specific MPI implementations and remains accessible without requiring privileged system setup.

\begin{figure*}[th]
    \centering
    \includegraphics[width=\textwidth]{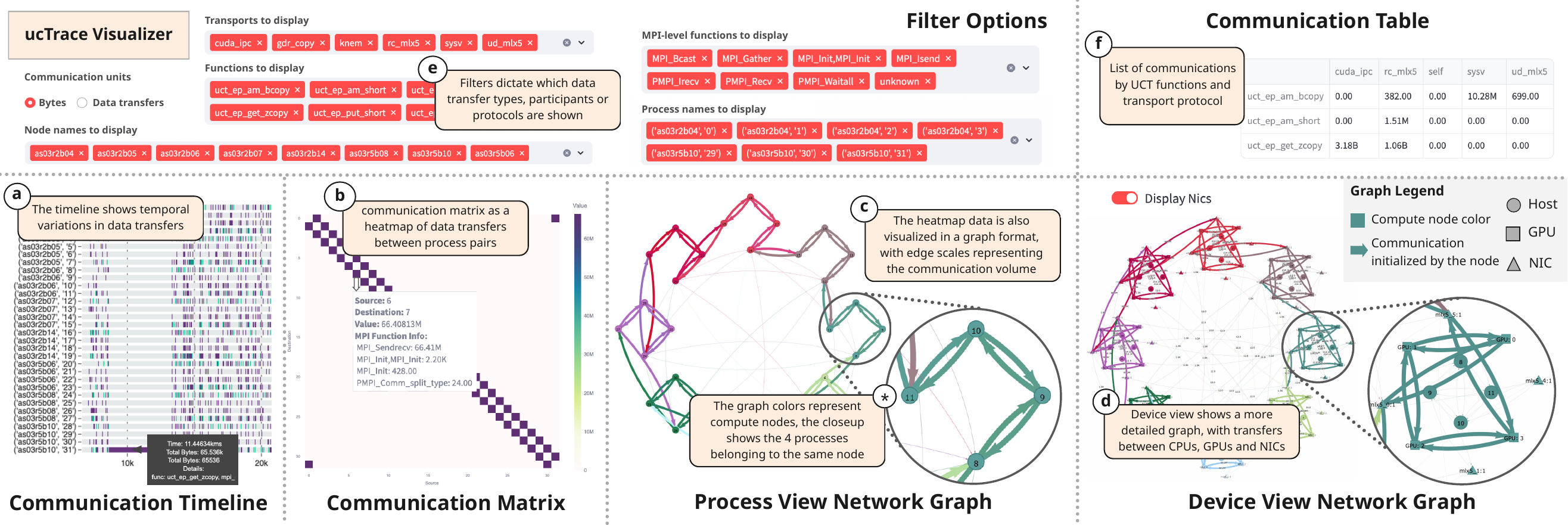} 
    \caption{Screenshots of the different features of \tool{} visualizer}
    \label{fig:tool-vis}
\end{figure*}
\section{Methodology}
\label{sec:methodology}

\tool{} intercepts and collects communication events at both the \uct{} and \ucp{} layers, linking \ucp{} {\tt send} and {\tt receive} operations and associating \uct{}-level transfers with their corresponding \ucp{} operations. It further attributes these communications to higher-level MPI functions and identifies the source and target GPU devices involved. As a result, \tool{} builds a comprehensive profile of \uct{} communications, enriched with GPU and NIC-level context, as well as MPI-layer associations. 

Figure \ref{fig:tool-flowchart} illustrates the workflow of \tool{}. A modestly modified \uct{} library is dynamically linked against the target executable to capture inline \uct{} {\tt send} function calls. When the user application makes a \ucx{} communication, a \uct{} primitive is called and that \uct{} communication is intercepted by \tool{} \textcircled{1}. For each \uct{} {\tt send}, \tool{} stores the {\tt send} function family, source \ep{} and \iface{} attributes such as memory domain name, transport layer name and the associated NIC. To collect timing information of these \uct{} communications, \tool{} replaces the completion callback of the \uct{} primitive with a wrapper function and records the time of the function call \textcircled{2}. To provide MPI rank information for visualization, \tool{} calls {\tt{MPI\_Comm\_rank}} upon MPI initialization and stores the rank \textcircled{3}. During the execution, the device attributor stores device memory allocations and frees from each process to support GPU attribution using Compute Sanitizer by NVIDIA \textcircled{4}. After the execution is complete, stored values from each process and device attributor outputs are analyzed and combined, and a python pickle file is generated that contains the merged and analyzed communications \textcircled{5}. The visualizer accepts the generated pickle file and creates interactive visuals \textcircled{6}.

\subsection{\tool{} Visualization}

Before diving into the details of the implementation of \tool{}, Figure \ref{fig:tool-vis} shows an overview of its visualizer. 
The visualizer shows data transfers on multiple levels of granularity, has a wide range of filtering options to configure the displayed data, allows tracking the communications in the time dimension and provides fine-grained details about individual data transfers, devices, communication types and protocols. 

The communications timeline, labeled \circled{a} on Figure \ref{fig:tool-vis}, shows the communications of each process over time, highlighting periods when no communication occurs. Hovering over a communication displays an information block with details about the communication. The communication matrix visualizes the communication between each process using a heatmap \circled{b}. Hovering over a cell reveals the distribution of MPI-level communications within that cell. The process view network graph \circled{c} displays the communication between host processes. 

The processes belonging to the same node are shown with the same color. The device view network graph \circled{d} separates this communication to GPUs, NICs, and hosts, which are represented as squares, triangles and circles, respectively. The arrows represent the logical communications between the illustrated devices and do not represent topological links. To display NICs, each \uct{} communication that has a network device associated to its \iface{} and remote \iface{} is split into three visualized communications. The three communications are from the source device to the source NIC, the source NIC to the target NIC and the target NIC to the target device. For both the communication matrix and the network graphs, the user can filter \circled{e} the displayed communications with respect to \uct{} protocols, \uct{} level functions or MPI level functions, as well as participating node or process. Moreover, the data can be visualized in terms of number of communications or bytes transferred. Lastly, a table \circled{f} shows the most frequently observed UCT functions with a breakdown of the corresponding transports. 

\edit{
    These visualization choices were driven by practical intuition and experience. The communication matrix \circled{b} and timeline \circled{a} are standard in performance analysis tools~\cite{vampir,paraver,pipit,Snoopie} and help users identify communication patterns and bottlenecks. The process view \circled{c} and device view \circled{d} network graphs build on top of Snoopie's system view~\cite{Snoopie}, extending it to support multi-node views and represent host processes, GPUs, and NICs on the same graph. The device view, in particular, was developed to leverage \tool{}'s fine-grained data-path tracing capabilities.
}

\subsection{Recording \uct{} Communications}
\uct{} supports different communication operations. These are \textit{short} for small messages, \textit{bcopy} for buffered copy and \textit{zcopy} for large messages with zero-copy operations. Moreover, each operation can utilize remote memory operations (\textit{get}, \textit{put}) and \textit{active messages}. Each operation and remote memory access combination is a different inline function that calls the provided \ep{}'s internal function in \uct{}. To intercept these functions, \tool{} makes minimal modifications to {\tt uct.h} and {\tt uct\_iface.c} files, adding an empty proxy function inside the inline functions. Once \tool{} intercepts the communication function, it records the current time, the source \ep{}, and other arguments provided to the {\tt send} function.

Endpoints of different transport protocols can support multiple \uct{} communication types. To understand the nature of the communication, \tool{} needs to extract information about the source and target devices, the transport protocol that is being used and which operation is used. Internally, the communication functions of \ucx{} call a function pointer in the \ep{} struct, which casts the \ep{} to its transport-specific version. To avoid implementing transport specific logic, \tool{} intercepts \iface{} creation, \ep{} creation, and \ep{} connection functions and infer the nature of the communication using this information.

Intercepting \iface{} creations provides \tool{} with transport information, the network devices used by the \iface{}, and optionally the \iface{} address. Similarly, intercepting the \ep{} creation function reveals the \ep{} address and, if available, the initial connection for that \ep{}. Finally, intercepting the \ep{} connection function provides information about established \ep{} connections.

 At the end of the execution, \tool{} creates a log file capturing all the information it has recorded: \iface{} information, \ep{} information, \ep{} connections  with timestamps, communications with timestamps.

\subsection{Tracking \uct{} Completions}
\label{sec:uct-completions}

\tool{} provides information on how long \uct{} communications take. Some \uct{} communications such as \textit{short} messages complete when the function returns a success flag. However, other communications including \textit{zcopy} are non-blocking\sout{do not complete when the function returns}. Instead, these functions return an in-progress flag and update a \textit{completion} provided by the user upon completion. These completions contain a counter and a callback function. When a completion is updated, its counter is reduced, and once the counter becomes zero, the callback is called. This is useful when one MPI communication is split into multiple \uct{} communications, and the callback is called when all the communications are completed.

There are two requirements for replacing the user provided completion with a custom one: (1) Even if multiple communications were originally sharing the same completion handle, \tool{} needs to give each communication a unique completion handle to track individual completion times. (2) After all communications sharing a completion handle are complete, the original completion function should be called to not break program behavior.
To satisfy these requirements, ideally, the replaced communication handle should store the original callback, which is not possible because C programming language does not allow creating anonymous functions and using them as function pointers.

To satisfy these requirements, \tool{} defines $N$ functions that have the same function signature of a completion callback, each with a unique ID from 1 to $N$. The original completion handles are stored in a map, indexed by the replacement's ID. Upon completion, the completion time is recorded, the original completion is updated and the original completion is only called when its count reaches 0. $N$ is defined during compile time, and the functions are generated by a Python script automatically. For our experiments, $N$ was set to 128, however in practice we observed that a much smaller number ($<50$) would suffice.

\subsection{UCP Attribution}
\label{subsec:ucp-attr}


Applications and libraries that support \ucx{} such as Open MPI and MPICH typically rely on \ucp{} for its high-level, user-friendly interface. Therefore, providing only \uct{}-level information may be of limited use to users interested in monitoring MPI.
Moreover, source and target address information cannot be obtained for active messages in \uct{} without protocol-specific logic. By associating \uct{} operations with higher-level \ucp{} operations, it becomes possible to retrieve source and target memory addresses for active messages and identify the MPI function that triggered the \ucp{} operation.

\ucp{} attribution is implemented by intercepting \ucp{} sends and receives. Currently, \tool{} only tracks tagged non-blocking \ucp{} operations as these are the only operations used by Open MPI's {\texttt {ucx pml}}. Thankfully, \ucp{} completion functions accept a custom user-provided argument. For each \ucp{} communication, \tool{} wraps the original user-provided argument with additional information, and replaces the original completion function with a custom one. When the custom completion callback is called, \tool{} records the \ucp{} communication and the original completion is called with the user-provided argument.

For each \ucp{} communication, the \uct{} \ep{}s managed by that \ucp{} \ep{}, the buffer, communication size, the call stack trace, starting and ending times are recorded. These values are written to the log files at the end of the execution.

\subsection{MPI Attribution}
MPI implementations such as Open MPI and MPICH supports multiple network transports including \ucx{}. \sout{While Open MPI supports a {\tt btl} (byte level transfer) transport that uses \uct{} directly, it is generally disabled during compile time.}
Open MPI can be configured to use \ucx{} for point to point (p2p) communications, OpenSHMEM support and RMA operations using \ucx{} {\tt pml} (p2p messaging layer), {\tt spml} (shared memory p2p messaging layer) and {\tt osc} (one-sided communication layer). 
These modules use \ucp{} level functions. Similarly, MPICH uses \ucp{} level functions when utilizing \ucx{} for communications.

\tool{} achieves MPI attribution by collecting call stack information for each \uct{} and \ucp{} operation. During log processing, these call stacks are examined to determine whether they contain an MPI function. \tool{} also intercepts {\tt MPI\_Init} calls, and executes {\tt MPI\_Comm\_rank} at exit to associate each process with its MPI rank.

\ucp{} attribution is necessary for MPI attribution because \ucp{} can schedule \uct{} messages asynchronously, hiding the originating MPI call from call stacks. Moreover, without \ucp{} attribution, remote RMA get operations resulting from \ucp{} rendezvous protocols (e.g., {\tt get\_zcopy}) appear to be initiated from {\tt MPI\_Wait} functions rather than the actual send operations.

\subsection{GPU Device Attribution}
\label{subsec:device-attr}

\ucp{} supports device pointers to be passed as payloads to communications and chooses the appropriate \uct{} transports to be used. Applications that use \ucx{} for communications such as MPI implementations like Open MPI and MPICH take advantage of this feature of \ucp{} to provide GPU-aware functionality such as CUDA-aware MPI and ROCm-aware MPI \cite{rocm-aware-MPI}. The GPU-aware property of MPI is widely used, hence, it is useful to differentiate GPU communications from regular communications and indicate source and target devices for GPU based communications. \tool{} attributes \ucx{} communications to source and target devices to provide information on GPU based communications.

\tool{} uses \sanitizer{}'s to record CUDA memory allocation and free operations with time information. Source and target pointers of \uct{} and \ucp{} communications are compared against device attributor logs to determine if the messages were sent from or to device memory. \uct{} active messages and buffered copy operations do not contain source or target buffer information, and therefore rely on \ucp{} attribution for correct device attribution. Active Messages \cite{active-messages} include the ID of a handler function which could be user defined, and buffered copy operations use a user-defined argument with a packing function to transfer data to the internal send buffer. As a result, it is not possible to obtain source or target memory addresses when \am{}s or buffered copies are used in \uct{}. By linking these \uct{} operations to their corresponding \ucp{} operations, \tool{} obtains the source and target device information that would otherwise be inaccessible at the \uct{} layer. 

\edit{
\tool{} currently only supports \nvidia{} device attribution but support for other GPU vendors is possible with a similar approach. Only the device attributor module of \tool{} depends on \sanitizer{}, and other modules are decoupled from it. Other GPU vendors provide similar software capable of recording memory allocation and free operations of device memory. For example AMD provides ROCprofiler SDK~\cite{rocprofiler-sdk} and Intel offers oneAPI Level Zero\cite{intel_level_zero_spec} which provides similar features. We could implement our device attributor with these tools for these GPU vendors, without modifying the main \tool{} code. 
}

\sout{
While \ep{} connections can identify source and target processes, they do not indicate whether the communication involves devices. For instance, a {\tt cuda\_ipc} communication provides source and target pointers without revealing their device associations. Similarly, a zero-copy communication in the {\tt ib} domain using {\tt rc\_mlx5} transport could represent either a GPU Direct RDMA or a host memory RDMA transfer.

One way to perform device attribution is by tracking two main components: memory management operations (i.e., memory allocations) and \uct{} send operations.
This enables associating a {\tt send} buffer with a source device if it falls within a previously recorded memory allocation range. For correctness, device memory {\tt free} operations must also be tracked, as virtual address space can be reused for different memory types. However, communicating device allocation information to remote peers would be required to identify whether a target address in a \uct{} {\tt send} is a device address. Instead, \tool{} records timing information with device memory allocations and logs them. These logs are later analyzed to determine source and target devices for all \uct{} and \ucp{} {\tt send} and {\tt receive} operations. This approach is preferred over using CUDA’s \texttt{cuPointerGetAttribute}, which is limited by failures on freed pointers and pointers allocated by other processes.
For correctness, we also need to keep track of device memory {\tt free} operations as the virtual address space can be reused for different types of memory. However this approach requires this device memory allocation information to be communicated to remote peers to determine if the target memory address is a device address in a \uct{} {\tt send} operation. Instead, \tool{} stores the time information alongside the device memory allocation operations, and logs them. These logs are analyzed at the time of log processing to determine source and target devices for all \uct{} and \ucp{} {\tt send} and {\tt receive} operations.This approach is preferred over relying on CUDA’s \texttt{cuPointerGetAttribute}, which has several limitations such as failing on freed pointers or pointers allocated by other processes.
}


\sout{
\subsubsection{\textbf{UCT Active Message Device Attribution}}
We discussed how device attribution is performed for \uct{} communications that provide source and target memory pointers. However, \uct{} active message communications do not expose either pointer.
Active Messages represent a communication paradigm in which a header is sent alongside the data. The header includes the address of a remote handler function, which the peer executes upon message arrival \cite{active-messages}. \ucx{} implements this mechanism by associating each active message with an ID transmitted along with the message.
The remote peer registers a valid \am{} handler to an \am{} ID. Since \uct{} allows user defined functions to be registered as \am{} handlers, it is not possible to get the target memory address from the {\tt send} operation. Moreover, the buffered-copy \am{}s do not take source memory address. Instead, these operations expect a user-defined argument and packing function that transfers the data to the given internal buffer.

By linking \uct{} \am{}s to \ucp{} operations, and applying device attribution to \ucp{} operations, \tool{} obtains the source and target devices for \uct{} \am{}s.}

\subsection{Log Processing}
Upon completion, each process generates two log files: one for device memory allocations (used for device attribution) and another for \uct{} and \ucp{} communication logs. 
 The {\em Log Processing} stage analyzes these files to perform four tasks:
1) link \uct{} communications to remote processes, 2) match \ucp{} sends with \ucp{} receives, 3) apply device attribution to \uct{} and \ucp{} communications, and 4) associate \uct{} communications to \ucp{} communications.


\textbf{1. Linking \uct{} communications to processes:} 
Recorded \uct{} communications include an address field, which can represent an \ep{}, \iface{}, or device address. For each communication, the most recent connection preceding it that uses the same \ep{} is identified. If the remote connection address is an \iface{} address, all remote \ep{}s are examined to find a match with the remote address. If the remote connection address is an \ep{} address, the logs of all processes are searched to find the most recent connection prior to the local connection that mirrors it.

\textbf{2. Device Attribution on \uct{} and \ucp{} communications:} 
For each \uct{} and \ucp{} communication, we compare the communication time and source/target pointers with device attribution logs to associate the addresses with specific devices.

\textbf{3. Matching \ucp{} sends with \ucp{} receives:}
When \ucx{} is built in debug mode, \ucp{} \ep{}s store the peer process ID. During {\em Log Processing}, the earliest {\tt receive} operation from the same peer ID that completes after the {\tt send} operation begins and has a matching tag is assigned to the {\tt send}. For intra-node communications, Open MPI uses a unique tag for each peer pair, and the earliest {\tt receive} operation is matched with the earliest {\tt send} operation, therefore, clock drift between nodes is not an issue.

\textbf{4. Associating \uct{} communications with \ucp{} communications:}  
For each \uct{} communication, candidate \ucp{} communications are first filtered such that the {\tt send} originates from the source process and the corresponding {\tt receive} belongs to the target process. The candidates are then narrowed down by verifying whether the \ucp{} send \ep{} manages the \uct{} \ep{}. Finally, the source and target buffers of the \uct{} communication are compared with those of the \ucp{} communication to determine a match.

When log processing is complete, a Python pickle file is generated that contains the curated logs for visualization.

\begin{figure*}[th]
    \centering
    \includegraphics[width=\textwidth]{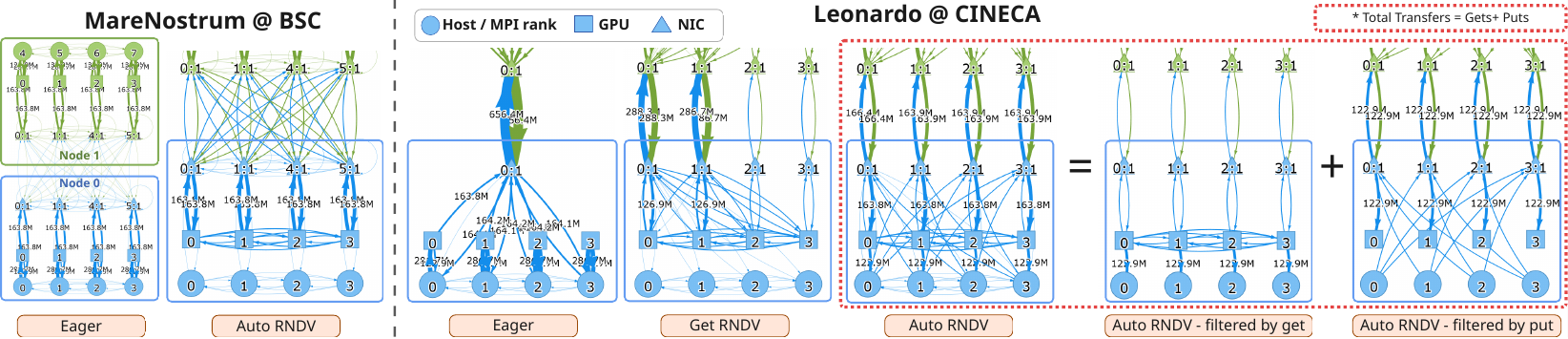} 
    \caption{CUDA-aware {\tt MPI\_ISend} with eager and different rendezvous protocols}
    \label{fig:ucx-rndv-fig}
\end{figure*}

\section{Evaluation}
\label{sec:evaluation}

In this section, we present a series of experiments to demonstrate the capabilities of \tool{}. 

\textbf{Hardware:} \tool{} is evaluated on the MareNostrum 5 supercomputer at the Barcelona Supercomputing Center unless stated otherwise and the CINECA's Leonardo supercomputer for experiment \ref{sec:ucx-rndv}. Each node in MareNostrum is equipped with two Intel Xeon Platinum 8460Y, four NVIDIA Hopper H100 GPUs and four ConnectX-7 NDR200 InfiniBand NICs. Leonardo is equipped with an Intel Xeon Platinum 8358, four NVIDIA Ampere GPUs and two dual port HDR100 per node. 

Note that GPU attribution is not currently supported for AMD systems, as the AMD machines available to us use HPE Cray Slingshot interconnects, which are not fully supported by \ucx{}, where OFI-based backends are preferred.
 
\textbf{Software:}
\tool{} was developed and tested with \ucx{} version {\tt 1.16.x}.
\tool{} is easily installable in user space, requiring only \ucx{} and {\tt nlohmann/json}. Applications that directly call \uct{} functions must be rebuilt with \tool{}, while those using \ucp{} operations (e.g., Open MPI, NVSHMEM) do not require rebuilding.

For all experiments, \ucx{} 1.16.0 compiled with CUDA and GDRCopy support was used. On MareNostrum, Open MPI 4.1.5 included in {\tt nvidia-hpc-sdk} 24.9 module was used for experiments in Sections \ref{sec:ucx-rndv} and \ref{sec:mpi-allreduce}. Open MPI 4.1.7a1 included in {\tt nvidia-hpc-sdk 24.3} module with {\tt hpcx 2.17.1} was used for the experiments in Sections \ref{sec:cg-experiment} and \ref{sec:gromacs}. 
MPICH 4.3.0 was compiled from source as the MPICH module on MareNostrum did not have CUDA support. On Leonardo, Open MPI 4.1.5 from the provided nvhpc@24.3 package from Spack was used.

\subsection{\ucx{} Eager and Rendezvous Protocols}
\label{sec:ucx-rndv}
\ucx{} supports \textit{eager} and \textit{rendezvous (rndv)} protocols for zero-copy operations. \textit{Eager} sends the data with no preceding notification while \textit{rndv} sends a notification first and completes the data transfer later. 
\textit{Rndv} supports two schemes: {\tt get\_zcopy} utilizes RDMA\_READ from the receiver while {\tt put\_zcopy} utilizes RDMA\_WRITE from the sender. 
By default the \textit{rndv} threshold and the \textit{rndv} scheme is automatically determined by \ucx{} to provide the best performance. 



\begin{figure*}[t]
    \centering
    \includegraphics[width=\textwidth]{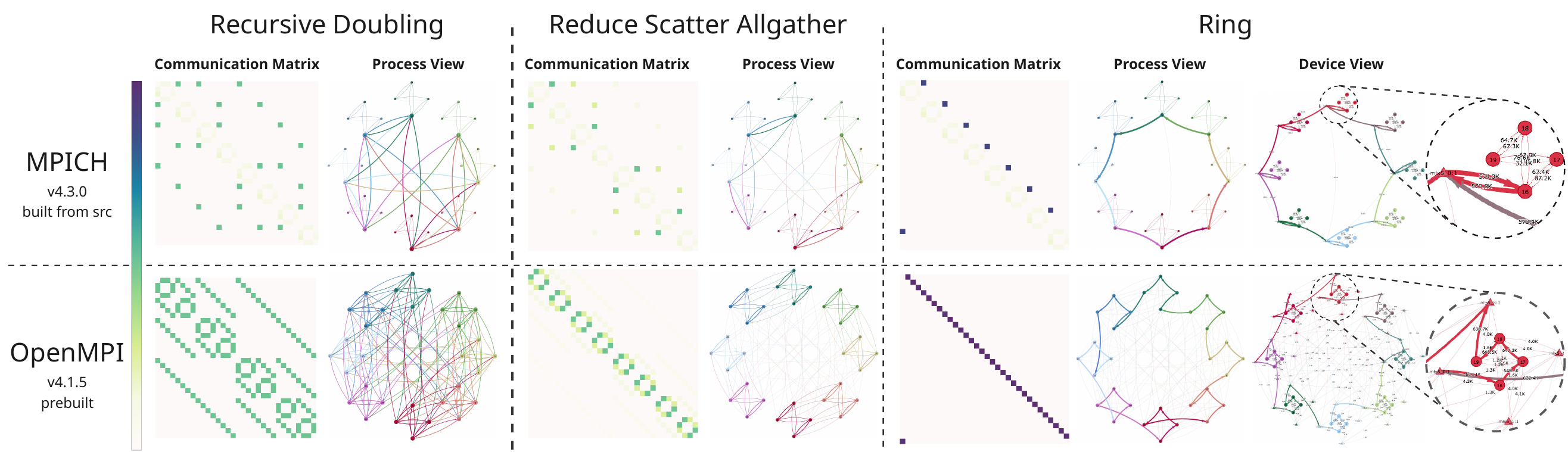} 
    \caption{Communication graphs and communication matrices for {\tt MPI\_Allreduce} on 8 nodes each with 4 cores run with recursive doubling, reduce scatter allgather and ring algorithms using Open MPI v4.1.5 and MPICH v4.3.0 on MareNostrum 5}
    \label{fig:collectives}
\end{figure*}

In this experiment, we use CUDA-aware Open MPI to send GPU buffers to all other MPI ranks using \texttt{MPI\_Isend}. We conduct runs with various UCX configurations to analyze \textit{eager} and \textit{rndv} protocols, as well as different \textit{rndv} schemes. Figure \ref{fig:ucx-rndv-fig} illustrates the communication patterns observed. These observations made by \tool{} can be valuable for both system admins and communication library developers. 

First of all, \ucx{} behaves differently on Leonardo and MareNostrum in terms of data paths. On MareNostrum, each GPU communicates with a dedicated NIC, and all NICs communicate with each other. In contrast, on Leonardo, GPUs communicate to different NICs, and each NIC communicates exclusively with a single corresponding remote NIC. 

For \textit{rndv} protocols, \ucx{} uses \texttt{cuda\_ipc} for intra-node GPU-to-GPU communication, while the eager protocol relies on \texttt{gdr\_copy} \cite{gdrcopy} along with \texttt{sysv}. For inter-node communication, eager uses an active message pattern: data is sent as an active message, and the \texttt{gdr\_copy} handler on the receiver side places the data on the target device. These behaviors can be observed using \tool's filters.

When set to automatic, \ucx{} exclusively uses the \textit{get rndv} on MareNostrum. In contrast, Leonardo employs both \texttt{put} and \texttt{get} schemes complementarily, as shown in Figure~\ref{fig:ucx-rndv-fig} using \tool{} filters. This is confirmed by forcing \ucx{} to use only \textit{get rndv}, which differs from the auto mode and is also shown in the figure as \textit{Get RNDV}.
In auto mode, \texttt{get\_zcopy} with \texttt{cuda\_ipc} handles intra-node GPU communication, while \texttt{cuda\_copy} moves data to the host. The host data is then written to the target  using \texttt{put\_zcopy} over \texttt{rc\_mlx5}. Notably, each GPU sends data directly to its NIC, so each host rank communicates only with the other three NICs, evident in the "\textit{auto RNDV filtered by put}" case in Figure~\ref{fig:ucx-rndv-fig}.

\subsection{Open MPI and MPICH AllReduce}
\label{sec:mpi-allreduce}
{\tt Allreduce} is a common collective with multiple algorithmic variants optimized for different system configurations and message sizes. MPI selects the best one dynamically at runtime. This experiment configures Open MPI and MPICH to use three specific {\tt AllReduce} algorithms: \textit{Recursive Doubling}, \textit{Reduce-Scatter-Allgather}, and \textit{Ring}. While both libraries support more variants, we chose these three due to their direct counterparts in both implementations.


We use the prebuilt Open MPI 4.1.5 on MareNostrum, but build MPICH 4.3.0 from source since the available version was not CUDA-aware. Consequently, some differences may arise from configuration or build environment. Our goal is not to compare the two MPI implementations, but to demonstrate \tool{}’s capabilities, showing how it can assist library developers or system administrators in building, configuring, and verifying library behavior.

Figure~\ref{fig:collectives} shows communication matrices and process-level graphs for different {\tt Allreduce} algorithms in Open MPI and MPICH, plus device-level graphs for the Ring algorithm. MPICH typically assigns one process per node for inter-node communication, while Open MPI involves multiple processes. This is evident in the first two algorithms: both MPI libraries communicate with the same nodes, but MPICH channels inter-node traffic through a single process (thicker arrows), whereas Open MPI distributes it (thinner arrows).

A similar pattern is seen in the Ring algorithm. In MPICH, the process-level communication graph forms a ring using one process per node as the communication point. In contrast, Open MPI involves all processes forming a ring. 
As a result, MPICH uses a single NIC for both sending and receiving, while Open MPI utilizes separate NICs, one for receiving, one for sending, clearly illustrated in the device view graph generated with \tool{}.

\subsection{Communication Graph Analysis of CG}
\label{sec:cg-experiment}
Conjugate Gradient (CG) solves large linear systems of the form $Ax = b$, where $A$ is sparse, symmetric, and positive definite. In this experiment, we profile a multi-GPU CG implementation~\cite{acg}. Each CG iteration involves an SpMV (Sparse Matrix-Vector multiply), where each GPU holds a partition of matrix $A$ and the part of input vector $x$. 
We use the METIS partitioner~\cite{karypis-kumar-1998} to distribute matrix rows by minimizing the edge cut.
To compute SpMV, GPUs exchange parts of $x$ corresponding to nonzero columns in their local matrices using {\tt MPI\_Isend} and {\tt MPI\_Irecv} primitives. 
For our experiments, we used two sparse matrices (Hook\_1498 and nd24k) from SuitSparse. 
While nd24k is a smaller matrix, it is denser than Hook\_1498 as shown in Table \ref{table:matrix-info}. 

\begin{table}[t]
\centering
\caption{Matrix size and nnzs for Hook\_1498 and nd24k}
\label{table:matrix-info}
\begin{tabular}{|l|r|r|r|r|}
\hline
\textbf{Matrix Name} & \textbf{Rows} & \textbf{NNZs} & \textbf{Mean NNZs/Rows}\\
\hline
Hook\_1498 & 1,498,023 & 59,374,451 & 39.6\\
nd24k & 72,000 & 28,715,634 & 398.8 \\
\hline
\end{tabular}
\end{table}

\begin{figure}[t]
    \centering
    \includegraphics[width=0.5\textwidth]{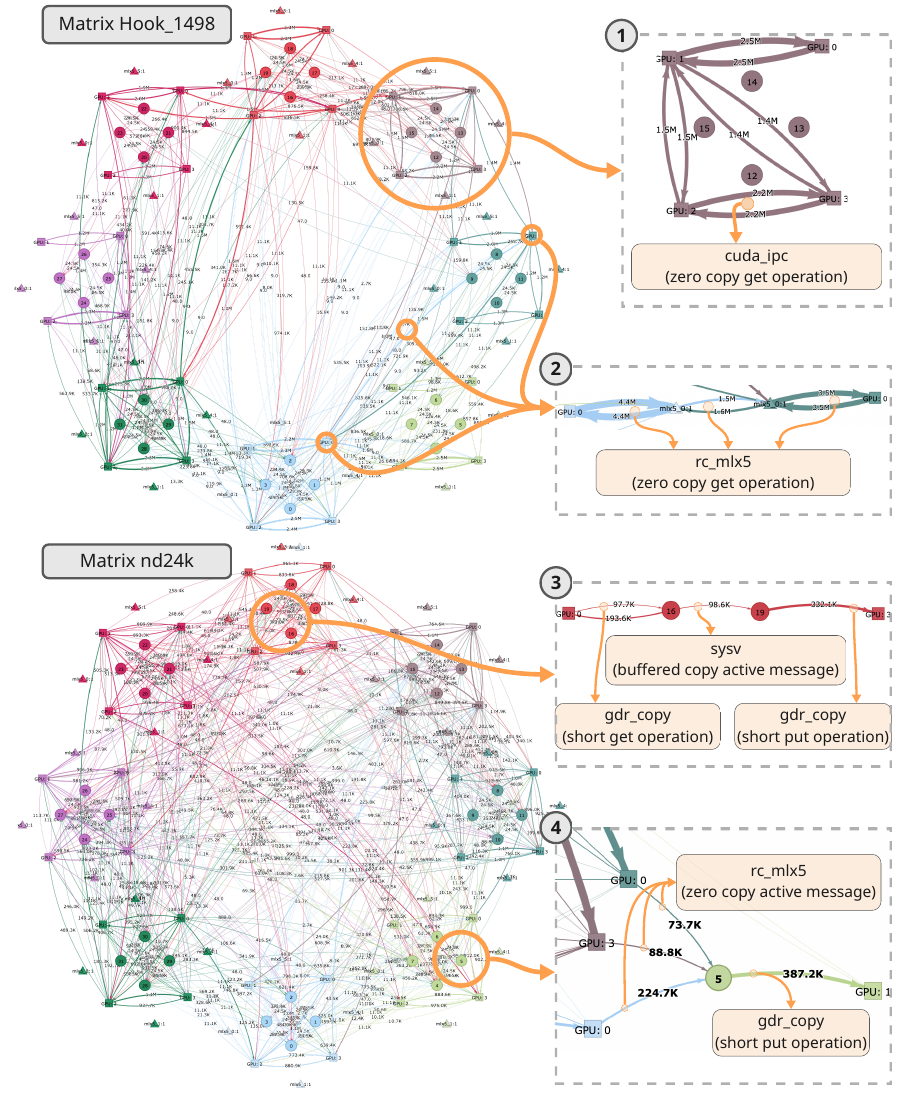} 
    \caption{Communication graphs of two sparse matrices filtered for p2p communications highlighting 4 commonly used GPU communication patterns by \ucx{}}
    \label{fig:cg-1}
\end{figure}

\newcommand{\ndk}{\texttt{nd24k}}
\newcommand{\hook}{\texttt{Hook\_1498}}

Figure \ref{fig:cg-1} displays communication graphs of  these two matrices used for solving CG with Open MPI on MareNostrum. 
Using the filters in \tool{}, the figure only displays {\tt {MPI\_Isend}} and {\tt {MPI\_Irecv}}. Because \hook{} has more than two times the number of nonzeros compared to \ndk{}, it requires more data to be transferred. However, \ndk{} is denser than \hook{}, and as a result the nonzero values in each GPU's partial matrix are distributed across more columns, requiring data transfer from/to more ranks. Conversely, \hook{} has larger message sizes  compared to \ndk{}'s smaller but more frequent messages. This communication behavior is illustrated in Figure \ref{fig:cg-1}, where \hook{}'s communication pattern appears as thicker but less frequent arrows, contrasting with \ndk{}'s thinner but more frequent arrows.


Figure \ref{fig:cg-1} also highlights four common communication patterns observed in solving CG. These patterns are \textcircled{1} zero copy get operation with {\tt {cuda\_ipc}} transport, \textcircled{2} zero copy get operation with {\tt {rc\_mlx5}} transport, \textcircled{3} buffered copy active messages with {\tt {gdr\_copy}} transport to copy data to/from the host buffer and {\tt sysv} transport to copy the buffer to peer process, and \textcircled{4} zero copy active messages with {\tt {rc\_mlx5}} transport to transfer data from GPU to remote host and {\tt {gdr\_copy}} transport to put the data to the remote GPU. 

Pattern \textcircled{1} uses {\tt cuda\_ipc} to transfer GPU buffers between GPUs on the same node. Since {\tt cuda\_ipc} leverages NVLINK for high bandwidth and low latency, \ucx{} uses it for most intra-node GPU-GPU communications. However, it requires sending a memory handle via the host, creating a bottleneck. For small messages, \ucx{} may choose other transports. For instance, \tool{} observed active messages via a host buffer, as in pattern \textcircled{3}, for intra-node communication with matrix \ndk{}.



Pattern \textcircled{2} uses {\tt {rc\_mlx5}} transport which forms a reliable connection using MLX5 drivers for NVIDIA NICs over InfiniBand to enable inter-node RDMA operations. As discussed in Section \ref{sec:ucx-rndv}, \ucx{} prefers the {\tt get} RNDV schema for large messages, and this pattern is frequently used for inter-node GPU to GPU communication for both matrices.

Pattern \textcircled{3} uses a three step approach for buffered copy. 
First, the {\tt {gdr\_copy}} transport is used to transfer the data to the host by utilizing GDRcopy \cite{gdrcopy}. 
Then, the {\tt {sysv}} transport is used to transfer the data to the target process via SystemV shared memory. Finally, another {\tt {gdr\_copy}} operation is used to put the data to the target device memory. 
Nevertheless, this pattern is rare, accounting for under $0.4\%$ of total data transfers, and only observed with \ndk{}.

\begin{table}[tp]
  \centering
  \scriptsize
  \caption{Top contributing transports and UCT functions filtered for P2P communications in CG}
  \label{table:top-contibutors-cg}
  {
  \begin{tabular}{|l|rrrr|r|}
    \hline
    \multicolumn{6}{|c|}{Bytes Transferred \% (Number of Data Transfers \%)}\\
    \hline
    \textbf{Hook\_1498} & \textbf{cuda\_ipc} & \textbf{gdr\_copy} & \textbf{rc\_mlx5} & \textbf{sysv} & \textbf{total} \\
    \hline
    am\_bcopy & 0.0 (0.0) & 0.0 (0.0) & 0.0 (0.1) & 0.9 (10.9) & 0.9 (11.0) \\
    am\_short & 0.0 (0.0) & 0.0 (0.0) & 0.5 (27.0) & 0.0 (10.9) & 0.5 (38.0) \\
    am\_zcopy & 0.0 (0.0) & 0.0 (0.0) & 2.5 (6.6) & 0.0 (0.0) & 2.5 (6.6) \\
    get\_short & 0.0 (0.0) & 0.0 (0.0) & 0.0 (0.0) & 0.0 (0.0) & 0.0 (0.0) \\
    get\_zcopy & 55.3 (10.9) & 0.0 (0.0) & 38.2 (26.7) & 0.0 (0.0) & 93.5 (37.6) \\
    put\_short & 0.0 (0.0) & 2.5 (6.6) & 0.0 (0.0) & 0.0 (0.0) & 2.5 (6.6) \\
    \hline
    total & 55.3 (10.9) & 2.5 (6.6) & 41.2 (60.4) & 0.9 (21.9) & 100.0 (100.0) \\
    \hline
    \textbf{nd24k} & \textbf{cuda\_ipc} & \textbf{gdr\_copy} & \textbf{rc\_mlx5} & \textbf{sysv} & \textbf{total} \\
    \hline
    am\_bcopy & 0.0 (0.0) & 0.0 (0.0) & 0.0 (0.1) & 1.7 (8.9) & 1.7 (9.0) \\
    am\_short & 0.0 (0.0) & 0.0 (0.0) & 0.9 (23.6) & 0.1 (8.2) & 1.0 (31.8) \\
    am\_zcopy & 0.0 (0.0) & 0.0 (0.0) & 8.8 (13.0) & 0.0 (0.0) & 8.8 (13.0) \\
    get\_short & 0.0 (0.0) & 0.4 (0.8) & 0.0 (0.0) & 0.0 (0.0) & 0.4 (0.8) \\
    get\_zcopy & 39.2 (8.2) & 0.0 (0.0) & 39.8 (23.4) & 0.0 (0.0) & 79.0 (31.5) \\
    put\_short & 0.0 (0.0) & 9.2 (13.8) & 0.0 (0.0) & 0.0 (0.0) & 9.2 (13.8) \\
    \hline
    total & 39.2 (8.2) & 9.5 (14.5) & 49.5 (60.1) & 1.8 (17.1) & 100.0 (100.0) \\
    \hline
  \end{tabular}
  }
\end{table}

Pattern \textcircled{4} uses {\tt {rc\_mlx5}} active messages with {\tt {gdr\_copy}} active message handlers to send inter-node eager messages as discussed in Section \ref{sec:ucx-rndv}. \ucx{} typically chooses to use active messages when the message size is small. As a result, this pattern is more common in \ndk{} compared to \hook{}. 

\subsection{Top Contenders Analysis in CG}

Next, we highlight the top contributors to communication, showcasing another useful feature of \tool{}. For clarity, we filter point-to-point communication in CG and present it as a table. 

Table~\ref{table:top-contibutors-cg} shows the percentage of bytes transferred and the number of data transfers for \uct{} communication primitives and transports. As discussed in Section~\ref{sec:cg-experiment}, patterns \textcircled{1} and \textcircled{3} enabled intra-node GPU-to-GPU communication. However, the {\tt gdr\_copy} transport with short get operations accounted for only $0.37\%$ of total bytes for \ndk{} and was unused for \hook{}, indicating that {\tt cuda\_ipc} was the preferred intra-node transport by \ucx{}. Notably, {\tt cuda\_ipc} accounted for $55.32\%$ and $39.15\%$ of total bytes for \hook{} and \ndk{}, respectively, but just $10.94\%$ and $8.16\%$ of the total transfers, indicating that thanks to the Metis partitioner, most communication remained within the node.


For inter-node communications, pattern \textcircled{2} was commonly observed, with zero-copy get operations using the {\tt rc\_mlx5} transport accounting for nearly $40\%$ of total bytes transferred for both matrices. Additionally, zero-copy active messages were used, contributing $2.5\%$ and $8.8\%$ of total bytes for \hook{} and \ndk{}, respectively. Pattern \textcircled{4} was used more frequently with the \ndk{} matrix, likely because its smaller size leads to smaller individual messages, where the RNDV protocol is less efficient.

The top contenders table also validates our pattern analysis from Section~\ref{sec:cg-experiment}. For pattern \textcircled{4} and matrix \hook{}, the percentages for bytes transferred and number of data transfers of {\tt am\_zcopy} with {\tt rc\_mlx5} match those of {\tt put\_short} with {\tt gdr\_copy}. For \ndk{}, this does not hold exactly, as {\tt gdr\_copy} is used in both patterns \textcircled{3} and \textcircled{4}. However, the difference in transfer counts between {\tt put\_short} with {\tt gdr\_copy} and {\tt am\_zcopy} with {\tt rc\_mlx5} ($13.76 - 13.00 = 0.76$) precisely matches the number of transfers for {\tt get\_short} with {\tt gdr\_copy}. 

\begin{figure}[t]
    \centering
    \includegraphics[width=.5\textwidth]{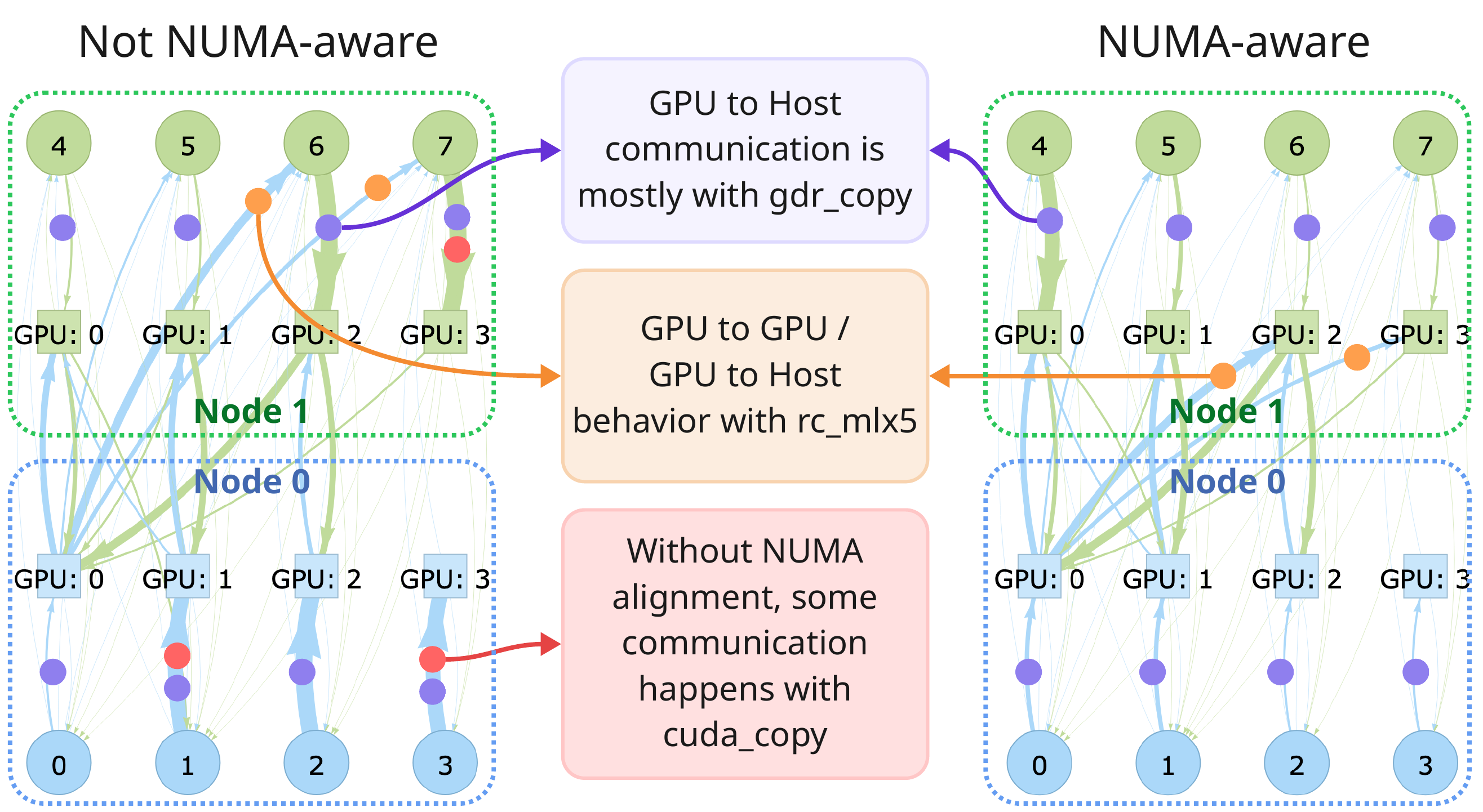} 
    \caption{CG communications with (right) and without (left) NUMA-awareness  filtered by nodes 0 and 1, displaying only {\tt {rc\_mlx5}}, {\tt {gdr\_copy}} and {\tt {cuda\_copy}} transports. Without NUMA-awareness, GPU-0 of node 0 communicates with host processes 6
and 7. With NUMA-awareness, these communications are made directly to GPU-2 and GPU-3 of node 1. 
    }
    \label{fig:numa-1}
\end{figure}



\subsection{Detecting a Performance Bug: NUMA
Affinity in CG}

In this experiment, we show how \tool{} helps application developers detect and diagnose initialization bugs that would otherwise go unnoticed, by exposing low-level communication behavior.

A NUMA (Non-Uniform Memory Access) node is a group of CPU cores that share local memory. In multi-GPU systems, each GPU typically connects to a specific NUMA node via PCIe, allowing faster transfers when accessed from its local domain. To ensure that each MPI rank uses a GPU local to its NUMA node, different binding strategies can be applied. One option is Open MPI’s \texttt{--bind-to numa}, which binds MPI tasks to NUMA nodes, but this alone does not ensure correct CPU-to-GPU affinity, as Open MPI lacks awareness of the system topology. Instead, we use a custom script to define CPU masks based on actual CPU–GPU mapping and pass it to \texttt{numactl} for precise control. In this case study, we compare CG solver performance on the \ndk{} matrix with and without explicit NUMA alignment.

Figure \ref{fig:numa-1} compares {\tt {rc\_mlx5}}, {\tt {gdr\_copy}} and {\tt {cuda\_copy}} communications of nodes 0 and 1. The communication graph on the left was obtained by  using {"\tt {--bind-to none}"}, allowing each MPI task in a node to be run on any CPU core while the graph on the right was obtained by explicitly setting the physical CPU mask according to the MPI rank using {\tt numactl}. Clearly, the non-NUMA-aware case cannot take advantage of GPU-GPU direct communications. Particularly, GPU-0 of node 0 communicates with host processes 6 and 7, and these host processes transfer the data to their respective GPUs. In contrast, in the NUMA-aware case, these communications, shown with orange circles, are made directly to GPU-2 and GPU-3 of node 1. 

In the non-NUMA-aware case, we observe intermittent use of the \texttt{cuda\_copy} transport for host-to-device transfers. It relies on \texttt{cudaMemcpyAsync} and is selected over \texttt{gdr\_copy} when memory is not registered or GPUDirect RDMA is unavailable. These \texttt{cuda\_copy} operations, highlighted in red, are absent in the NUMA-aware case. Without NUMA alignment, NIC utilization is suboptimal, only one NIC per node is used during \texttt{Allreduce}, versus four in the NUMA-aware case, making execution about five times slower.

\begin{figure*}[t]
    \centering
    \includegraphics[width=.95\textwidth]{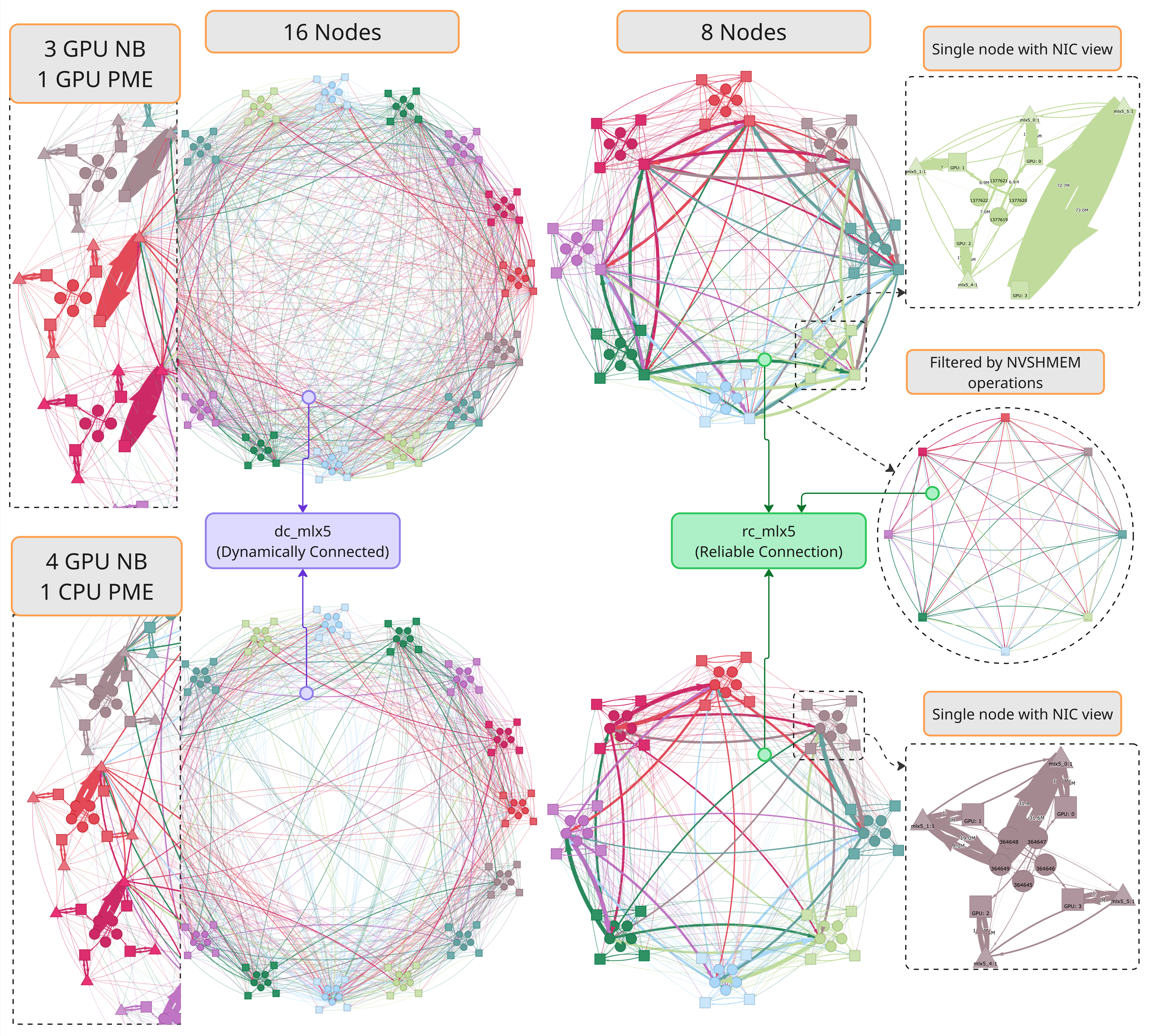} 
    \caption{GROMACS MD communications on 16 nodes (left) and 8 nodes (right) $\times$ 4 MPI tasks with three NB GPU tasks and one PME GPU task on each node (top), 16 nodes (left) and 8 nodes (right) $\times$ 5 MPI tasks with four NB GPU tasks and one PME host task on each node (bottom). NICs were hidden, GPUs and host tasks were kept visible. Segments of the figures were scaled up with NICs visualized. Low volume communications caused by {\tt hcoll} initialization were hidden. 8 node PME GPU figure was extended with filtered NVSHMEM operations.
    }
    \label{fig:gromacs}
\end{figure*}

\subsection{GROMACS Simulation}
\label{sec:gromacs}

GROMACS is a molecular dynamics (MD) simulation package widely used across scientific domains \cite{gromacs}, optimized for parallel execution with GPU-aware MPI.  We profiled GROMACS using {\tt water-cut1.0\_GMX50\_bare/0024} for 200 iterations with default parameters. The simulation consists mainly of two computational tasks: long-range electrostatic interactions via the Particle Mesh Ewald (PME) method, and short-range non-bonded (NB) interactions. We tested four configurations across 8 and 16 nodes: NB on 3 GPUs with PME on 1 GPU per node, and NB on 4 GPUs with PME on 1 CPU task. Running PME on multiple GPUs requires the {\tt cuFFTMp} library, which internally uses NVSHMEM. NVSHMEM was configured to use \ucx{} for communication; however, due to a known {\tt hpcx} bug \cite{NVIDIA_bug_fixes}, {\tt cuda\_ipc} transport was disabled. CUDA VMM support was also disabled since \tool{}'s device attributor does not support that API.

Figure~\ref{fig:gromacs} displays the profiled communications for this experiment. In all four experiments, \ucx{} uses RDMA over InfiniBand for intra-node GPU-GPU communication because {\tt cuda\_ipc} transport was disabled. This can be seen in the figures where the NICs are visualized as there are no direct arrows between GPUs, instead arrows are seen between NICs in the same node.

Each GPU uses a distinct NIC for inter-node communication and each host task uses a single NIC, except for the host-side PME experiment on 8 nodes, where PME host tasks use two NICs. From the figures, most inter-node communication originates from the PME GPUs (top) or PME host tasks (bottom). \tool{} logs confirm that in all four experiments, $69\%$–$75\%$ of inter-node communication was caused by PME tasks. Consequently, most inter-node communication passes through a single NIC per node.

For the 16-node experiment with host-side PME, $26\%$ of all communication consisted of {\tt MPI\_Alltoall} calls, which are primarily FFT transpose operations, while $10\%$ comprised {\tt MPI\_Sendrecv} calls from 
the grid summation function, and $8\%$ were 
calls for coordinate exchange. The remaining $50\%$ of communication was attributed to GPU halo exchange for NB computations. 
For the device-side PME configuration, $30\%$ of all communication involved 
GPU halo exchanges using {\tt MPI\_Isend}. An additional $14\%$ consisted of NVSHMEM communications conducted internally by {\tt cuFFTMp} for all-to-all FFT transpose operations, while $40\%$ was due to NB GPU halo exchange operations. Figure~\ref{fig:gromacs} includes a filtered view of the all-to-all NVSHMEM operations in the 8-node experiment.

While \ucx{} used {\tt rc\_mlx5} transport for inter-node RDMA operations in all the previous experiments, on 16 nodes, \ucx{} uses {\tt dc\_mlx5} transport for such operations. This transport leverages Dynamically Connected (DC) Queue Pairs in the MLX5 driver for scalable, reliable communications. \ucx{} selects the appropriate inter-node communication transport based on system topology and communication characteristics: RC (Reliable Connection)  is preferred for fewer tasks due to its lower overhead, while DC is used for better scalability with more tasks \cite{rc-vs-dc}. 

\subsection{Overhead}

The time and memory overhead of \tool{} is shown in Table \ref{table:overhead}. The overhead data was collected on MareNostrum 5, on the GROMACS and CG experiments. The GROMACS experiment, was run for 20,000 steps with PME tasks on the host. The CG experiment was run on the \texttt{nd24k} matrix with 1000 iterations. As the table suggests, capturing the call-stack information significantly increases the overhead. \edit{ \tool{} currently uses \texttt{glibc}'s \texttt{backtrace} to collect call-stack information for each \uct{} and \ucp{} call, which is the primary source of overhead. One approach for reducing this overhead would be using a more efficient backtrace approach such as \texttt{libunwind}'s \texttt{unw\_backtrace} instead of \texttt{backtrace}, which has the same API as \texttt{backtrace}, but is significantly faster~\cite{backtrace}. 

Another approach of reducing call-stack recording overhead is recording entry and exit of relevant functions such as \texttt{PMPI} family of functions and associating the intercepted \ucx{} calls to that function. This method would allow MPI-level attribution avoiding full call-stack captures. While these lighter-weight approaches would substantially reduce overhead, we leave them for future work.} Overall, we find \tool{}’s runtime overhead and log size reasonable for use in real-world applications.

\begin{table}[tbp]
  \centering
  \scriptsize 
  \setlength{\tabcolsep}{4pt} 
  \caption{\tool{} Performance and resource usage comparison for GROMACS and CG benchmarks across a varying number of GPUs. The table shows baseline performance versus \tool{} tracing overhead, along with the resulting log size and memory consumption.}
  \label{table:overhead}
  \begin{tabular}{
    S[table-format=3.0, table-space-text-post=\textsuperscript{*}]
    S[table-format=2.3] 
    S[table-format=3.3] 
    S[table-format=2.1] 
    S[table-format=2.3] 
    S[table-format=2.3] 
  }
    \toprule
    {\textbf{\#GPUs}} & {\textbf{Baseline(s)}} & {\textbf{ucTrace(s)}} & {\textbf{Overhead(x)}} & {\textbf{Log(GiB)}} & {\textbf{Mem(GiB)}} \\
    \midrule
    \multicolumn{6}{c}{\textit{-- GROMACS --}} \\
    \midrule
    16                  & 19.976 & 26.156   & 1.3  & 0.663  & 4.700  \\
    16\textsuperscript{*} & 19.976 & 160.984  & 8.0  & 1.300  & 7.000  \\
    32                  & 12.623 & 24.466   & 1.9  & 1.700  & 5.100  \\
    32\textsuperscript{*} & 12.623 & 184.200  & 14.6 & 3.100  & 7.500  \\
    64                  & 10.811 & 30.713   & 2.8  & 3.200  & 4.800  \\
    64\textsuperscript{*} & 10.811 & 197.551  & 18.3 & 6.300  & 7.400  \\
    128                 & 9.409  & 59.413   & 6.3  & 9.200  & 7.000  \\
    128\textsuperscript{*}& 9.409  & 239.018  & 25.4 & 17.000 & 11.700 \\
    \midrule
    \multicolumn{6}{c}{\textit{-- CG --}} \\
    \midrule
    16                  & 0.236 & 0.567 & 2.4 & 0.040 & 0.092 \\
    16\textsuperscript{*} & 0.236 & 2.313 & 9.8 & 0.064 & 0.130 \\
    \bottomrule
    \multicolumn{6}{l}{\textsuperscript{*}\footnotesize{With call-stack.}}
  \end{tabular}
\end{table}

\section{Conclusion}
\label{sec:conclusion}
We introduced \tool{}, a multi-node profiling and visualization tool for analyzing \ucx{} communication patterns in distributed systems. \tool{} provides a multi-layer view of communication, enabling targeted analysis for system administrators, library developers, and application developers. Through detailed case studies, we demonstrated its utility in identifying performance bottlenecks, validating system configurations, and optimizing communication in both CPU- and GPU-based MPI applications. Our tool offers an effective solution for improving communication efficiency in modern HPC environments. Future work will extend \tool{}'s capabilities to support additional communication libraries to further broaden its applicability.

\section*{Acknowledgment}
The work is supported from the European
Research Council (ERC) under the European Union’s Horizon 2020 research and innovation programme (grant agreement No 949587).
We acknowledge EuroHPC Joint Undertaking for awarding access to the MareNostrum5 supercomputer in Spain and the Leonardo supercomputer in Italy (Project ID: EHPC-DEV-2024D10-091).

\bibliographystyle{IEEEtran}
\bibliography{references}

\end{document}